\documentstyle[12pt]{article}
\begin{document}
\begin{flushright}
hep-ph/9801349\\
Alberta Thy-02-98\\January, 1998\\
revised: March, 1998\\
revised: June, 1998
\end{flushright}
\vskip 1cm

\begin{center}
{\Large \bf $\pi \Lambda$ scattering phase shifts and $CP$ violation in $\Xi \rightarrow \pi \Lambda$ decay}\\[10mm]
A. N. Kamal \\[5mm]
{\em Theoretical Physics Institute and Department of Physics,\\ University of Alberta,
Edmonton, Alberta T6G 2J1, Canada.}
\end{center}
\vskip1cm

\begin{abstract}
$CP$-violating signals in weak $ \Xi\rightarrow \pi \Lambda$  decay require the knowledge of  $\pi \Lambda$ $S$- and $P$-wave  scattering phases at ${m}_{\Xi}$ center-of-mass energy. We have calculated these phases in baryon chiral perturbation theory with the ground-state $\Sigma$ in $s$ and $u$ channels and ${{3 \over 2}}^{+}$  $\Sigma(1385)$ in the $u$ channel. We do not treat the baryons as heavy. We find ${\delta}_{S} = {1.2}^{\circ}$ and ${\delta}_{P} = -{1.7}^{\circ}$ with the central value of the strong coupling parameter $D$. We also investigate the variation of the scattering phases as functions of the parameter $D$. We compare this result with previous calculations, and discuss its relevance to $CP$-asymmetry parameters.

\end{abstract}
PACS categories: 13.75.Gx,13.30.Eg,

\newpage

\begin{center}
{\bf I. INTRODUCTION}
\end{center} 

The only evidence to date for $CP$ violation in the six-quark model is found in second order ${K}^{0}-\bar{{K}^{0}}$ mixing. It is anticipated that direct (first order) $CP$ violation will be observed in $B$ decays. Though the potential for observing direct $CP$ violation also exists in hyperon weak decays, the field remains less well studied. The sign and the size of $CP$-violating signals in hyperon two-body hadronic decays depend on the difference between the  strong interaction phases of the $S$- and $P$-wave amplitudes in $\Lambda\rightarrow\pi N$, $\Sigma\rightarrow\pi N$, and $\Xi\rightarrow\pi \Lambda$ decays. For decays involving a pion and a nucleon in the final state, extensive low-energy phase shift analyses exist \cite{rop}. However, for $\Xi\rightarrow\pi \Lambda$ decay, one has to rely on the theoretical estimates of $\pi\Lambda$ $S$- and $P$-wave scattering phases until reliable measurements become available from the semileptonic decay  $\Xi\rightarrow \Lambda\pi e\nu$. Martin \cite{mar}, using a dispersive approach, calculates the $P$- wave phase shift ${\delta}_{P}$ to be $ \approx { -1.0}^{\circ}$. Nath and Kumar \cite{nath}, using perturbative tree graphs as the input in a $N/D$ method, estimate ${\delta}_{S} = {-18.7}^{\circ}$ and ${\delta}_{P} = {-2.7}^{\circ}$. Lu, Wise, and Savage \cite{lu}, using tree diagrams and chiral ${\mbox{SU}(2)}_{L}\times{\mbox{SU}(2)}_{R}$ in a heavy-baryon formulation of Ref. \cite{jen}, calculate ${\delta}_{S} = {0}^{\circ}$ and ${\delta}_{P} = {-1.7}^{\circ}$.  Subsequently, Datta and Pakvasa \cite{datta} extended the calculation of Ref. \cite{lu} by including the contributions of ${{ 1\over 2}}^{-}$ and ${{3 \over 2}}^{-}$ intermediate states in a heavy-baryon formulation of chiral ${\mbox{SU}(2)}_{L}\times{\mbox{SU}(2)}_{R}$, and found that the $S$-wave phase shift remains small, bounded by ${0.5}^{\circ}$. The authors of Ref. \cite{nath} obviously disagree with those of  Refs. \cite{lu} and \cite{datta} on the $S$-wave phase shift. The agreement among Refs. [3-5] on the $P$-wave phase shift is deceptive since its numerical value is so small. However, significantly, all  three calculations agree on the sign of the phase shift.

Considering that the baryons in the decay $\Xi\rightarrow\pi\Lambda$ do not involve heavy quarks, we have calculated the $S$ and $P$ wave phase shifts in a chiral ${\mbox{SU}(3)}_{L}\times{\mbox{SU}(3)}_{R}$ approach involving an octet of light spin-1/2 baryons \cite{kaymak}. The calculated phase shifts ${\delta}_{S}$ and ${\delta}_{P}$ for $\pi\Lambda$ scattering at $\Xi$ mass are in agreement with those of Lu, Wise and Savage \cite{lu} for the central value of the strong coupling parameter $D$. Thus for ${\delta}_{S}$ we disagree with Nath and Kumar \cite{nath}. This implies that the $CP$-violating signals will be quite small as implied in Ref. \cite{lu}.

We describe the method and analysis in Sec. II. The results are discussed in Sec. III.

\begin{center}
{\bf II. METHOD AND CALCULATIONS}
\end{center}

	The matrix element for a generic hyperon weak decay of  the kind ${B}_{i}({{1 \over 2}}^{+}) \rightarrow  {B}_{f}({{1 \over 2}}^{+}) \pi$ is expressed in terms of $S$- and $P$-wave amplitudes as
\begin{equation}
A({B}_{i} \rightarrow {B}_{f} \pi)  \propto S + P \vec{\sigma}\cdot\vec{q}.
\end{equation}
The total rate, for normalization purposes, is given by
\begin{equation}
\Gamma = {{G}_{F}}^{2}{{m}_{\pi}}^{4}{|\vec{q}|({E}_{f}+{m}_{f}) \over 4\pi{m}_{i}} ({|S|}^{2} + {|P|}^{2}).
\end{equation}

The angular distribution is proportional to \cite{com}
\begin{equation}
{d\Gamma \over d\Omega} \propto 1 + \gamma \vec{{\omega}_{i}}\cdot\vec{{\omega}_{f}} + (1 - \gamma) \hat{q}\cdot\vec{{\omega}_{i}} \hat{q}\cdot\vec{{\omega}_{f}} + \alpha\hat{q}\cdot(\vec{{\omega}_{i}}+\vec{{\omega}_{f}}) + \beta\hat{q}\cdot(\vec{{\omega}_{f}} \times \vec{{\omega}_{i}}),
\end{equation}
where $\vec{{\omega}_{i}}$ and  $\vec{{\omega}_{f}}$ are unit vectors in the direction of the initial and final spins, respectively.
The parameters $\alpha$, $\beta$ and $\gamma$, are given by
\begin{eqnarray}
\alpha = {2\mbox {Re} ({S}^{*} P) \over {{|S|}^{2} + {|P|}^{2}}}, \nonumber \\
\beta = {2\mbox{Im}({S}^{*} P) \over {{|S|}^{2} + {|P|}^{2}} }, \nonumber \\
\gamma = { {{|S|}^{2} - {|P|}^{2}} \over  {{|S|}^{2} + {|P|}^{2}}}.
\end{eqnarray}
The parameter $\alpha$ controls the decay asymmetry in the angular distribution if the final-state polarization is not measured:
\begin{equation}
{d\Gamma \over d\Omega} = {{G}_{F}}^{2} {{m}_{\pi}}^{4}{|\vec{q}|({E}_{f} + {m}_{f}) \over 16{\pi}^{2} {m}_{i}} ({|S|}^{2} + {|P|}^{2}) (1 + \alpha \hat{q}\cdot\vec{{\omega}_{i}}) .
\end{equation}

If $CP$ symmetry were respected, then \cite{com}
\begin{equation}
\bar{\Gamma} = \Gamma,~~~ \bar{\alpha} = - \alpha, ~~~ \bar{\beta} = -\beta, ~~~ \bar{\gamma} =\gamma.
\end{equation}

Invoking $CPT$ invariance, the decay amplitudes for the decay and their $CP$ conjugates are parametrized as follows:
\begin{eqnarray}
S = |S| {e}^{i({\delta}_{S}+{\phi}_{S})} , ~~~ P = |P| {e}^{i({\delta}_{P} + {\phi}_{P})}, \nonumber \\
\bar{S} = - |S| {e}^{i({\delta}_{S} - {\phi}_{S})} , ~~~ \bar{P} = |P| {e}^{i({\delta}_{P} - {\phi}_{P})} ,
\end{eqnarray}
where ${\delta}_{S,P}$ are the strong phases and ${\phi}_{S,P}$ the weak phases.

The $CP$-violating asymmetry parameter  ${\sl A}$ is then given by,
\begin{equation}
{\sl A} \equiv {\alpha + \bar{\alpha} \over \alpha - \bar{\alpha}} = - \mbox{tan}({\delta}_{S} - {\delta}_{P})    \mbox{tan}({\phi}_{S} - {\phi}_{P}) .
\end{equation}

\noindent Obviously, the size of the asymmetry parameter $\sl A$ depends on the difference ${\delta}_{S} - {\delta}_{P}$. We calculate the two phases ${\delta}_{S}$ and ${\delta}_{P}$ in baryon chiral perturbation theory. We do not treat the baryons as heavy as they do not involve heavy quarks.

	The chiral lagrangian involving the ${0}^{-}$ Goldstone-boson field $\phi$ and the ${{1 \over 2}}^{+}$ baryon field $B$ is given by \cite{kaymak}
\begin{eqnarray}
L_1= {{{f}_{\pi}}^{2} \over 8} \mbox{Tr} ({\partial}_{\mu}\Sigma {\partial}^{\mu}{\Sigma}^{\dagger}) + i \mbox{Tr} ( \bar{B}{\gamma}^{\mu}{\partial}_{\mu} B) - m \mbox{Tr} (\bar{B}B) \nonumber \\ + {i \over 2} \mbox{Tr} \bar{B} {\gamma}_{\mu} [\xi {\partial}^{\mu} {\xi}^{\dagger} + {\xi}^{\dagger} {\partial}^{\mu} \xi  ,B] \nonumber \\ + i {D \over 2} \mbox{Tr} \bar{B}{\gamma}_{\mu} {\gamma}_{5} \{ \xi {\partial}^{\mu} {\xi}^{\dagger} - {\xi}^{\dagger} {\partial}^{\mu} \xi , B\} \nonumber \\ + i {F \over 2} \mbox{Tr} \bar{B} {\gamma}_{\mu} {\gamma}_{5} [\xi {\partial}^{\mu} {\xi}^{\dagger} - {\xi}^{\dagger} {\partial}^{\mu} \xi , B],
\end{eqnarray}

\noindent with ${f}_{\pi} =131$ MeV, $D = 0.8 \pm 0.14$, $F = 0.5 \pm 0.12$ \cite{kaymak}, and 
\begin{equation}
\Sigma = \mbox{exp}(2i{M \over {f}_{\pi}}),
\end{equation}

\noindent where ${M}$ , ${B}$, and ${\bar{B}}$ are the standard ${{0}^{-}}$ meson and ${{1 \over 2}^{+}}$ baryon (antibaryon) octets \cite{gasio}. Under ${{\mbox {SU}(3)_L} \times{\mbox{SU}(3)_R}}$, 
\begin{eqnarray}
\Sigma\rightarrow L\Sigma {R}^{\dagger}, \nonumber \\
\xi \rightarrow L\xi {U}^{\dagger} = U\xi {R}^{\dagger}, \nonumber \\
{\xi}^{\dagger} \rightarrow U{\xi}^{\dagger} {L}^{\dagger} = R {\xi}^{\dagger} {U}^{\dagger}, \nonumber \\
B \rightarrow U B {U}^{\dagger}.
\end{eqnarray}

Expanding ${\xi}$ and ${{\xi}^{\dagger}}$ in powers of ${M}$ one can work out ${\pi\Sigma\Lambda}$ couplings. We note two things about this part of the calculation: (i) the linear ${\pi\Sigma\Lambda}$ coupling is of $D$ type only and (ii) there are no contact vertices of the kind ${(\pi\pi\Lambda\Lambda)}$. The latter is due to the fact that the quadratic terms in the pion field arise from the combination ${(\xi {\partial}^{\mu} {\xi}^{\dagger} + {\xi}^{\dagger} {\partial}^{\mu} \xi)}$ which results in an antisymmetric quadratic term in the pion field of the form $({\pi}_{1}{\partial}_{\mu}{\pi}_{2} - {\pi}_{2}{\partial}_{\mu}{\pi}_{1})$. Because of Bose statistics, this antisymmetric Lorentz form has to go with an antisymmetric isospin structure, namely, isospin 1. However, ${I = 1}$ exchange is not permitted in ${\pi\Lambda}$ scattering.

The relevant interaction of the ${{3 \over 2}^{+}}$ decuplet ${\Sigma(1385)}$ (denoted here by ${{\Sigma}^{\ast} }$) is introduced as follows:
\begin{equation}
L_2 = g{\bar{\Sigma}}_{\mu}^{\ast(a)} {\partial}^{\mu} {\pi}^{(a)}
\Lambda.
\end{equation}

The coupling constant $g$ is determined from the total width of ${\Sigma(1385)}$ and its branching ratio to ${\Lambda\pi}$ \cite{pdg} to be
\begin{equation}
{{g}^{2} \over 4\pi} = 7.03~{\mbox{GeV}}^{-2}.
\end{equation}

Let us now introduce the essentials of ${0}^{-}$-${1 \over 2}^{+}$ scattering formalism. The $T$ matrix is defined in terms of the amplitudes $A(s,t)$ and $B(s,t)$ as follows [11.12]:
\begin{equation}
T(s,t) = A(s,t) +{1 \over 2} \gamma({k}_{1} +{k}_{2}) B(s,t),
\end{equation}
where $s$ and $t$ are the Mandelstam variables; ${k}_{1}$ and ${k}_{2}$ are the pion momenta.

The non-spin-flip and spin-flip amplitudes ${f}_{1}(x)$ and ${f}_{2}(x)$, respectively, $x = \cos\theta$, where ${\theta}$ is the center-of-mass scattering angle, are defined in terms of $A(s,t)$ and $B(s,t)$ as follows [11,12]:
\begin{eqnarray}
{f}_{1} = {(E + m) \over8\pi\sqrt{s} } [ A(s,t) + (\sqrt{s} - m) B(s,t) ], \nonumber \\
{f}_{2} =  {(E - m) \over8\pi\sqrt{s} } [ - A(s,t) + (\sqrt{s} + m) B(s,t) ],
\end{eqnarray}
where $E$ is the baryon center-of-mass energy.

The partial waves ${f}_{L\pm}$ are projected out as follows \cite{pil},
\begin{equation}
{f}_{L\pm} = {1 \over 2} \int_{-1}^{1}{[ {P}_{L}(x) {f}_{1}(x) + {P}_{L\pm1} {f}_{2}(x) ]}.
\end{equation}

If we expand ${f}_{1}$ and ${f}_{2}$ in terms of partial waves
\begin{equation}
{f}_{1,2}(x) = \sum{ (2L + 1) {{f}_{1,2}}^{L} {P}_{L}(x)} ,
\end{equation}
then the partial waves we need, ${f}_{0+}$ and ${f}_{1-}$, are given by
\begin{eqnarray}
{f}_{0+} \equiv {f}_{S} = {{f}_{1}}^{L=0} + {{f}_{2}}^{L=1}, \nonumber \\
{f}_{1-} \equiv {f}_{P} = {{f}_{1}}^{L=1} + {{f}_{2}}^{L=0} .
\end{eqnarray}

Finally, the phase shifts are related to ${f}_{S}$ and ${f}_{P}$ by
\begin{equation}
{f}_{S,P} = {1 \over k} {e}^{i {\delta}_{S,P}} \sin{\delta}_{S,P},
\end{equation}
where $k$ is the center-of-mass momentum. It follows from Eq. (19) that,
\begin{equation}
k\cot{{\delta}_{S,P}} = \mbox{Re}  {[{f}_{S,P}]}^{-1}, ~~~~~~~~ k = - \mbox{Im} {[{f}_{S,P}]}^{-1}
\end{equation}
The second of Eq. (20) is the statement of elastic unitarity.

Consider now the process $\pi(k_1) + \Lambda(p_1) \rightarrow \pi(k_2) + \Lambda(p_2)$ in the center-of-mass frame. See Fig. 1. With the definitions of the Mandelstam variables
\begin{eqnarray}
s = (k_1 + p_1)^2, & t = (k_2 - k_1)^2, & u = (p_2 - k_1)^2 \nonumber \\
\mbox {with} & s + t + u = 2 {{m}_{\pi}}^{2} + 2 {{m}_{\Lambda}}^{2},
\end{eqnarray}
we find the following contributions to $A(s,t)$ and $B(s,t)$ arising from the spin-${1 \over 2}^{+}$ $\Sigma(1190)$ poles in the $s$ and $u$ channels:
\begin{equation}
{A}_{\Sigma}(s,t) = {({2D \over \sqrt{6}{f}_{\pi}})}^{2} ({m}_{\Lambda} + {m}_{\Sigma}) [ 2 + ({{m}_{\Sigma}}^{2} - {{m}_{\Lambda}}^{2}) \{{1 \over s-{{m}_{\Sigma}}^{2}} + {1 \over u-{{m}_{\Sigma}}^{2}}\}],
\end{equation}
\begin{equation}
{B}_{\Sigma}(s,t) =  {({2D \over \sqrt{6}{f}_{\pi}})}^{2} {({m}_{\Lambda} + {m}_{\Sigma})}^{2} [ {1 \over u-{{m}_{\Sigma}}^{2}} - {1 \over s-{{m}_{\Sigma}}^{2}}] .
\end{equation}

The contribution to $A(s,t)$ and $B(s,t)$ from the spin-${{3 \over 2}}^{+}$ $\Sigma(1385)$ (denoted here by ${\Sigma}^{*}$) in the $u$ channel is
\begin{eqnarray}
{A}_{{\Sigma}^{*}}(s,t) = - {g}^{2} ({m}_{{\Sigma}^{*}} +{m}_{\Lambda}) \{ {1 \over 3} + {t \over 2(u-{{m}_{{\Sigma}^{*}}^{2})}} + {({{m}_{{\Sigma}^{*}}}^{2} - {{m}_{\Lambda}}^{2}) \over 3(u-{{m}_{{\Sigma}^{*}}}^{2})}  \nonumber \\
 + {{{m}_{\Lambda}(u+{{m}_{{\Sigma}^{*}}}^{2} - 2{{m}_{\Lambda}}^{2})} \over {6{{m}_{{\Sigma}^{*}}}^{2}({{m}_{\Sigma}}^{*} + {m}_{\Lambda})}}  \nonumber \\
+ {{m}_{\Lambda} \over {6}{{m}_{{\Sigma}^{*}}}^{2}}{({{m}_{{\Sigma}^{*}}}^{2} - {{m}_{\Lambda}}^{2})({m}_{{\Sigma}^{*}} - {m}_{\Lambda}) \over u - {{m}_{{\Sigma}^{*}}}^{2}} \},
\end{eqnarray}
\begin{eqnarray}
{B}_{{\Sigma}^{*}}(s,t) = - 2 {g}^{2} \{{{m}_{\Lambda} \over 3}{({m}_{\Lambda}+ {m}_{{\Sigma}^{*}}) \over (u - {{m}_{{\Sigma}^{*}}}^{2})} - {t \over 4(u - {{m}_{{\Sigma}^{*}}}^{2})}  \nonumber \\
- {{m}_{\Lambda} \over 6 {m}_{{\Sigma}^{*}}} - {1 \over 12 {{m}_{{\Sigma}^{*}}}^{2}} (u + {{m}_{{\Sigma}^{*}}}^{2} - 2{{m}_{\Lambda}}^{2})  \nonumber \\
- {1 \over 12 {{m}_{{\Sigma}^{*}}}^{2}} {{({{m}_{{\Sigma}^{*}}}^{2} - {{m}_{\Lambda}}^{2})}^{2} \over u - {{m}_{{\Sigma}^{*}}}^{2}} - {{m}_{\Lambda} \over 6 {m}_{{\Sigma}^{*}}}{({{m}_{{\Sigma}^{ *}}}^{2} - {{m}_{\Lambda}}^{2}) \over (u - {{m}_{{\Sigma}^{*}}}^{2})}\}.
\end{eqnarray} 
 As all our amplitudes are real, the first of Eq. (20) implies that
\begin{equation}
{1 \over k} \tan{\delta}_{S,P} = {f}_{S,P},
\end{equation}
 where $k$ is the center-of-mass momentum. Note that as amplitudes calculated at the tree-level do not satisfy unitarity, the second of Eq. (20) is not satisfied.

 The projection of partial waves and the evaluation of the phase shifts is now straightforward. The following features of our calculation are worth noting. First and foremost,  because of an almost-complete cancellation between the contributions from ${A}^{L=0}$ and ${B}^{L=0}$ in Eq. (15), ${{f}_{1}}^{L=0}$ is very small; in fact, ${{f}_{1}}^{L=0}$ and ${{f}_{1}}^{L=1}$ are comparable for any given vlue of $D$. The smallest partial-wave amplitude is the spin-flip amplitude ${{f}_{2}}^{L=1}$. For the experimental range of $D$, ${{f}_{2}}^{L=0}$ is the largest partial-wave amplitude. Next, the contribution of ${\Sigma}^{*}$ in the $u$ channel to ${A(s,t)}^{L=0,1}$ and ${B(s,t)}^{L=0,1}$ is significant compared to that of $\Sigma$. In Table I we have tabulated the value of ${\delta}_{S}$ and ${\delta}_{P}$ as functions of $D$. As for the $P$-wave phase shift, it remains negative and small in the allowed range of $D$. For the central value of  $D = 0.80$, and $g$  from Eq. (13), we obtain at $\sqrt{s} = {m}_{\Xi}$ 
\begin{equation}
{\delta}_{S} = {1.2}^{\circ},  ~~~~ \mbox {and} ~~~~{\delta}_{P} = - {1.7}^{\circ}
\end{equation}

\vskip 5mm

\begin{center}
{\bf III. SUMMARY AND CONCLUSION}
\end{center}

	The $S$- and $P$-wave phase shifts calculated here  are consistent with those calculated in Refs. [5,7] and disagree with those of Ref. [4]. The depend on the parameter $D$. The $CP$-asymmetry parameter $\sl A$ would, therefore, be small as suggested in \cite{lu}.

The fact that we obtain ${\delta}_{S} \sim {1}^{\circ}$ for the allowed range of $D$, and Ref. [5] calculates ${\delta}_{S} = {0}^{\circ}$ is not too significant. In our calculation the smallness of ${\delta}_{S}$ results from an almost-complete cancellation between two relatively large numbers. (In contrast, the smallness of ${\delta}_{P}$ is due to small individual contributions.) The vanishing of ${\delta}_{S}$ as calculated in Ref. [5] can be understood as follows: In the heavy-baryon approximation adopted in [5,7], the baryon propagators are simplified according to the following replacements:
\begin{eqnarray}
s\mbox { channel:}~~~~{1 \over \gamma({p}_{1} +{k}_{1}) - {m}_{\Sigma}}~~\rightarrow~~{1 \over {m}_{\Lambda} + {E}_{\pi} - {m}_{\Sigma}} \nonumber \\
u \mbox {channel:}~~~~{1 \over \gamma({p}_{1} -{k}_{2}) -{m}_{\Sigma}}~~\rightarrow~~{1\over {m}_{\Lambda} - {E}_{\pi} - {m}_{\Sigma}}.
\end{eqnarray}

As a consequence, the scattering-angle dependence arising from the $u$-channel propagators is lost. The $\pi\Lambda$ scattering becomes, in effect, zero range. All scattering-angle dependence, now arising from the vertices only, becomes of finite-order-polynomial form in $\cos{\theta}$. Because of the derivative coupling of the pion field, the numerator of the $T$ matrix takes the following form for the $\Sigma$ intermediate-state diagrams:
\begin{equation}
S\cdot{k}_{1}S\cdot{k}_{2} = {1 \over 4}(v\cdot{k}_{1}  v\cdot{k}_{2} -{k}_{1}\cdot{k}_{2}) ={1 \over 4}({{k}_{1}}^{0}{{k}_{2}}^{0} -{k}_{1}\cdot{k}_{2}) ={1 \over 4}\vec{{k}_{1}}\cdot\vec{{k}_{2}},
\end{equation}
 where $S$ is the spin operator \cite{jen}. The structure in Eq. (29) evidently gives rise to $P$-wave scattering only. The $S$-wave amplitude vanishes strictly.
The ${\Sigma}^{*}$ intermediate-state diagram also generates only $P$-wave amplitude because the numerator of the spin-${3 \over 2}$ propagator in the heavy-baryon limit reduces to the form ${\delta}_{ij}$, where $i$ and $j$ are space-like indices. Thus, in the heavy-baryon limit $S$-wave scattering amplitude vanishes, and $P$-wave is the only other partial wave generated. We emphasize that the vanishing of the $S$-wave amplitude is not simply due to the derivative coupling, but the derivative coupling {\it and} the heavy-baryon approximation. This vanishing of the amplitude occurs individually for $\Sigma$- and ${\Sigma}^{*}$-exchange amplitudes.

 \vskip 5mm 
\begin{center}
{\bf Acknowledgments}
\end{center}

 An individual operating grant from the Natural Sciences and Engineering Research Council of Canada  partially supported this research. I wish to thank Dr. Alakabha Datta for pointing out numerical errors in an earlier version of this paper.
\newpage
\begin{table}
\begin{center}
\caption {Variation of the phase shifts with the parameter $D$. Phase shifts are expressed in degrees.}
\vskip 3mm

\begin{tabular}{|c|c|c|c|c|}
\hline

Phases&$D=0.6$&$D=0.7$&$D=0.8$&$D=0.9$ \\
\hline
$\delta_{S}$&$0.86$&$1.03$&$1.23$&$1.72$ \\
\hline
$\delta_{P}$&$-0.50$&$-1.07$&$-1.72$&$-2.47$ \\
\hline
\end{tabular}
\end{center}
\end{table}

\begin{figure}

%%Begin InstantTeX Picture
\let\picnaturalsize=N
\def\picsize{5.0in}
\def\picfilename{fig1.eps}
%If you do not have the picture file add:
%\let\nopictures=Y
%to the beginning of the file.
\ifx\nopictures Y\else{\ifx\epsfloaded Y\else\input epsf \fi
\let\epsfloaded=Y
\centerline{\ifx\picnaturalsize N\epsfxsize \picsize\fi \epsfbox{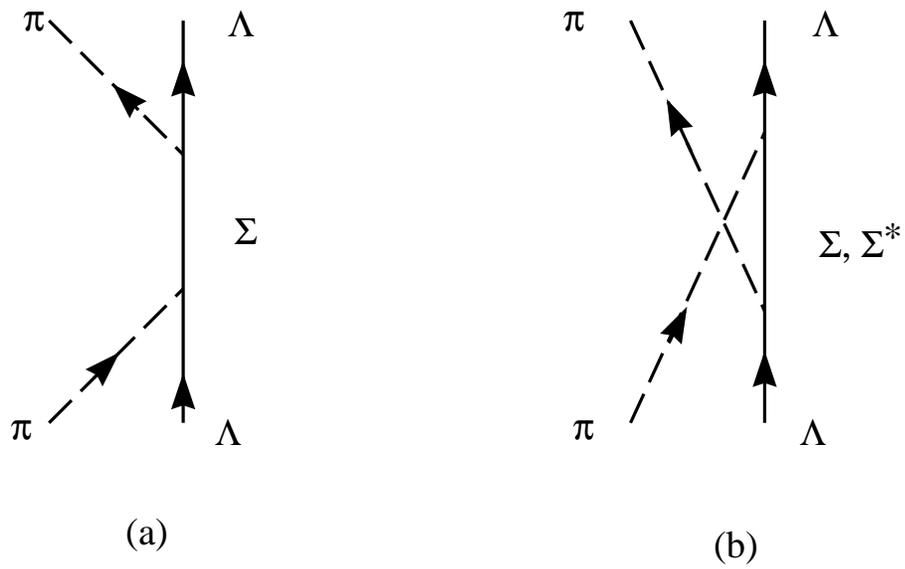}}}\fi
%%End InstantTeX Picture

\caption {$\pi \Lambda$ scattering. (a): $s$-channel diagram. (b): $u$-channel diagram.}
\end{figure}


\newpage
\begin{thebibliography}{99}
\bibitem{com}
E. Commins and P. Bucksbaum, {\sl Weak Interactions of Leptons and Quarks} (Cambridge University Press, New York, 1983).
\bibitem{rop}
See, for example, L. D. Roper, R. M. Wright, and B. T. Feld, Phys. Rev. {\bf 138}, B190 (1965); J. Hamilton and W. S. Woolcock, Rev. Mod. Phys. {\bf 35}, 737 (1963); A. Donnachie, J. Hamilton, and A. T. Lea, Phys. Rev. {\bf 135}, B515 (1964).
\bibitem{mar}
B. R. Martin, Phys. Rev. {\bf 138}, B1136 (1965).
\bibitem{nath}
R. Nath and A. Kumar, Nuovo Cimento {\bf 36}, 669 (1965).
\bibitem{lu}
M. Lu, M. B. Wise, and M. J. Savage, Phys. Lett. B {\bf  337}, 133 (1994).
\bibitem{jen}
E. Jenkins and A. V. Manohar, Phys. Lett. B {\bf  255}, 558 (1991).
\bibitem{datta}
A. Datta and S. Pakvasa, Phys. Lett. B {\bf  344}, 430 (1995).
\bibitem{kaymak}
See, for example, \"{O}. Kaymakcalan, Lo Chong-Huah and K. C. Wali, Phys. Rev. D {\bf  29}, 1962 (1984); Xiao-Gang He and G. Valencia, Phys. Lett. B {\bf  409}, 469 (1997).
\bibitem{gasio}
See, for example, S. Gasiorowicz, {\sl Elementary Particle Physics} (Wiley, New York ,1967), Chap. 18.
\bibitem{pdg}
Particle Data Group, R. M. Barnett {\it et al.}, Phys. Rev. D {\bf  54}, 1 (1996).
\bibitem{chew}
G. F. Chew, M. L. Goldberger, F. E. Low, and Y. Nambu, Phys. Rev {\bf 106}, 1337 (1957).
\bibitem{pil}
H. Pilkuhn, {\sl The Interaction of Hadrons} (Noth-Holland, Amsterdam, 1967).
 

\end{thebibliography}
\end{document}